\documentclass[10pt,twocolumn]{revtex4}
\usepackage{graphicx} 
\usepackage{dcolumn}
\usepackage{bm}
\pdfoutput=1
\usepackage{amssymb} 
\usepackage{amsmath}
\usepackage{color}

\newcommand{\BE}{\begin{equation}}
\newcommand{\EE}{\end{equation}}
\newcommand{\BA}{\begin{eqnarray}}
\newcommand{\EA}{\end{eqnarray}}

\newcommand{\erf}{{\mbox{erf}}}

\begin{document}

\title{Optimal run-and-tumble based transportation of a Janus particle with active steering}
%\shorttitle{Title} %Insert here a short version of the title if it exceeds 80 characters

\author{Tomoyuki Mano}
\affiliation{Department of Physics, The University of Tokyo, Tokyo, Japan}

\author{Jean-Baptiste Delfau}
\affiliation{IFISC (CSIC-UIB), Instituto de F\'{\i}sica
Interdisciplinar y Sistemas Complejos, E-07122 Palma de
Mallorca, Spain}

\author{Masaki Sano}
\affiliation{Department of Physics, The University of Tokyo, Tokyo, Japan $\,$}

\date{\today}

%abstract - max 800 characters
%
\begin{abstract}
Even though making artificial micrometric swimmers has been made possible by using various propulsion mechanisms, guiding their motion in the presence of thermal fluctuations still remains a great challenge. Such a task is essential in biological systems, which present a number of intriguing solutions that are robust against noisy environmental conditions as well as variability in individual genetic makeup. Using synthetic Janus particles driven by an electric field, we present a feedback-based particle guiding method, quite analogous to the ``run-and-tumbling'' behavior of \textit{Escherichia coli} but with a deterministic steering in the tumbling phase: the particle is set to the “run” state when its orientation vector aligns with the target, while the transition to the ``steering'' state is triggered when it exceeds a tolerance angle $\alpha$. The active and deterministic reorientation of the particle is achieved by a characteristic rotational motion that can be switched on and off by modulating the AC frequency of the electric field, first reported in this work. Relying on numerical simulations and analytical results, we show that this feedback algorithm can be optimized by tuning the tolerance angle $\alpha$. The optimal resetting angle depends on signal to noise ratio in the steering state, and it is demonstrated in the experiment. Proposed method is simple and robust for targeting, despite variability in self-propelling speeds and angular velocities of individual particles.
\end{abstract}

\maketitle

%%%%%%%%%%%%%%%%%%%%%%%%%%%%%%%%%%%%%%%%%%%%%%%%%

The physics of active suspensions made significant progress during the past decades and it is now possible to build artificial microscopic particles able to self-propel in a fluid. The range of possible applications of such swimmers is wide, with fascinating perspectives: targeted drug delivery\cite{Sundararajan08}, bottom-up assembly of very small structures\cite{Ismagilov02}, mixing or automatic pumping in microfluidic devices\cite{Paxton06}, design of new microsensors and microactuators in MEMS\cite{Fennimore03} or artificial chemotactic systems\cite{Howse07} to name a few. A lot of man-made microscopic swimmers fall into the category of ``Janus'' particles which share the same property: an asymmetric structure inducing a breaking of symmetry of the interactions with the surrounding fluid resulting in a self-propelling force. Several physical phenomena can be at the origin of this force: local temperature gradients induced by a defocused laser beam\cite{Jiang10} (thermophoresis), enzimatic catalysis of chemical reactions by a coated surface\cite{Paxton06,Howse07,Fournier-Bidoz2005,Golestanian05,Paxton04} or electrostatic interactions between surface charges and the ions of the solution\cite{Gangwal08} (induced-charge electrophoresis or ICEP).\\
If several methods are known to generate self-propelling forces for Janus particles, guiding their motion remains a challenge. The biggest difficulty consists in controlling their orientation, a particularly delicate task when working with microscopic objects subjected to thermal fluctuations. Swimmers need to resist rotational diffusion by fixing or steering their orientation to reach specified targets or follow given trajectories. Experimental works showed that it was possible to lock the orientation of catalytic nanorods made of ferromagnetic materials using magnetic fields\cite{Kline05}. Another interesting method involves visualizing the orientation of the particle at every moment and turn on the self-propelling force only when it is directed to the right direction\cite{Qian13}. In that approach, the reorientation process is ``passive'' in a sense that the experimentalist waits for rotational diffusion to correct the orientation of the particle.\\
In this paper, we use Janus particles driven by ICEP and introduce a new method to control their trajectory with an ``active'' reorientation process. This new concept consists in switching between two distinct modes of motion exhibited by the particles: a self-propelling state and a regular rotation state. Such rotations had already been observed experimentally with L-shaped self-propelling swimmers moving by thermophoresis\cite{Kummel13} but the origin and characteristics of these rotations are very different here. The Janus particle under feedback control exhibits a motion quite similar to the ``run-and-tumbling'' behavior observed for the bacteria \textit{Escherichia coli}\cite{Berg72}. However, the reorientation is not random but deterministic, which might be compared to the adaptive steering found in evolved organisms, e.g. phototaxis in \textit{Volvox carteri}\cite{Drescher10}. Such a ``hybrid'' strategy enables a high efficiency while minimizing the complexity of the implementation. In the first part of this article, we will describe in detail the two different behaviors exhibited by our Janus particles. Based on these properties, the second part will be devoted to the experimental implementation of the proposed particle guiding method. Finally  in the third part, we will present numerical simulations and analytical calculations showing how it is possible to optimize the feedback process.

\begin{figure}
\begin{tabular}{cc}
\includegraphics[width=8 cm]{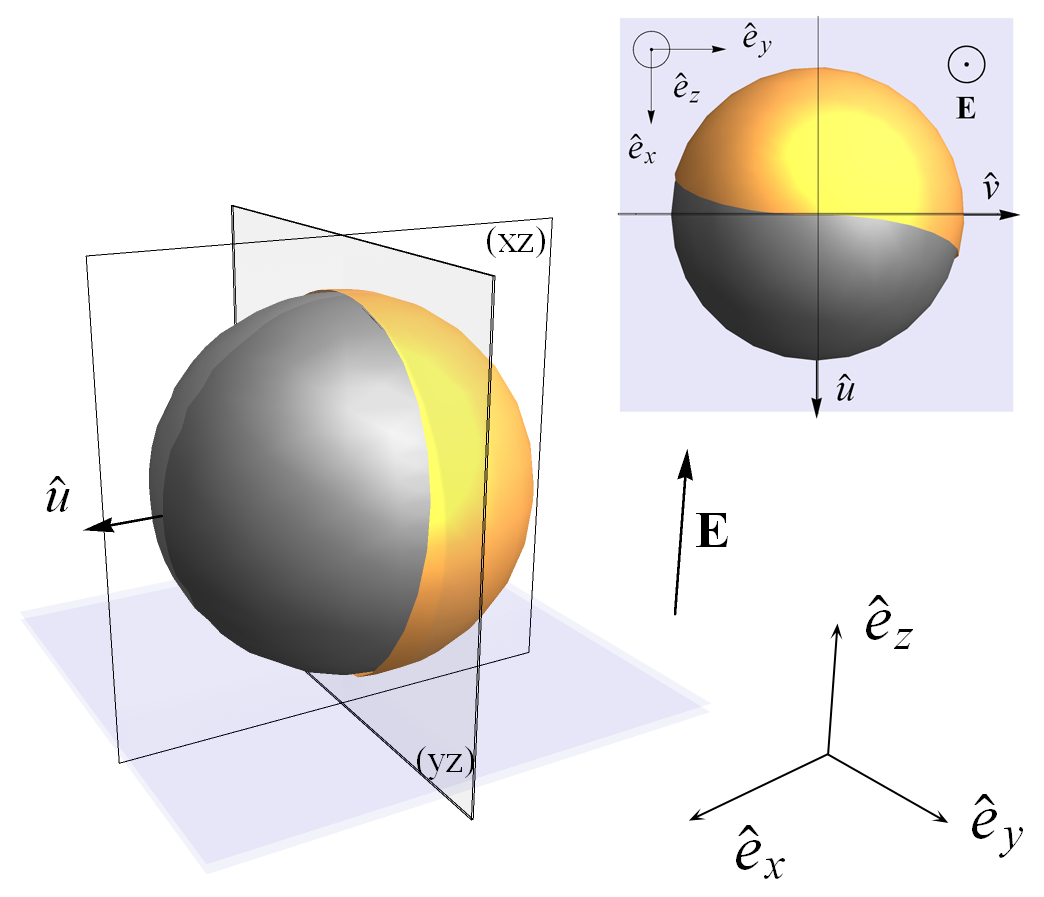}
\end{tabular}
\caption{\label{scheme_Janus}Three-dimensional scheme of a Janus particle with an electric field $\boldsymbol{E}$ parallel to $\boldsymbol{\hat{e}_z}$. Inset: top view of a chiral Janus particle slightly asymmetric with respect to the $(xz)$ plane}
\end{figure}

\section{Individual behavior of Janus particles}

Our experimental device is described in detail in the materials and methods section. It is very similar to the one first introduced by Gangwal et al\cite{Gangwal08}: the self-propelling motion of the particles is obtained by applying an AC electric field $\boldsymbol{E} = E_0/2 \sin(\omega t) \, \boldsymbol{\hat{e}_z}$ to the solution (see Fig.~\ref{scheme_Janus}). We can control the peak-to-peak voltage $E_0$ and the frequency $\omega$ of the electric field so that we have two control parameters. Janus particles can exhibit very interesting individual and collective behaviors depending on the values of $\left( \omega, E_0 \right)$ (see~\cite{Nishiguchi15,Suzuki11} for a detailed description). In this work, we will focus on two particular individual types of motion: ``active Brownian motion'' (ABM) and rotations.

\subsection{Active Brownian Motion (ABM)}

As soon as $\boldsymbol{E}$ is applied to the solution, the particles are attracted by the electrodes and restoring forces lock the cross section across the equator between two hemispheres parallel to the electric field ($\boldsymbol{\hat{e}_z}$ axis)\cite{Kilic11}. The combined effects of gravity and electric forces transport the particles close to the bottom electrode. For low frequencies (roughly between $500 \mbox{ Hz}$ and $30 \mbox{ kHz}$), Janus particles self-propel at a constant speed $U_0$ in the direction of the dielectric hemisphere\cite{Gangwal08}: this is the Self-Propelling region (SP). For higher frequencies ($\omega > 30 \mbox{ kHz}$), the direction of propagation is reversed and the particles move in the direction of the metal side: this is the Inverse Self-Propelling region (ISP)\cite{Suzuki11}. In both cases, their motion is 2-dimensional in the plane (xy) perpendicular to the electric field (see Fig.~\ref{scheme_Janus}). The direction and amplitude of their velocity depend on the two control parameters $(\omega, E_0)$. For lower frequencies, $U_0$ basically increases when $E_0$ increases or $\omega$ decreases.\\

The self-propulsion mechanism in the SP region is well understood in the framework of Induced-Charge Electro-Phoresis (ICEP)\cite{Squires04, Squires06}: the Induced-Charge Electro-Osmotic (ICEO) flow around the particle - resulting from the electro-osmotic flow of counter ions in double layers on the metal and dielectric hemispheres - is asymmetric because of the different polarizabilities of the hemispheres. ICEO fluid flow induces a constant self-propelling force $\boldsymbol{F}$ acting on the particles in the direction of the dielectric hemisphere as well as restoring forces preventing them from rotating around $\hat{e}_x$ or $\hat{e}_y$. The ICEP theory predicts that $F \propto E_0^2$. On the other hand, the origin of $\boldsymbol{F}$ in the ISP region still lacks a theoretical explanation. In the SP or ISP regions, the particles exhibit an ``Active Brownian motion'': as they are subjected to rotational diffusion around the $\boldsymbol{\hat{e}_z}$ axis, their motion will be diffusive at long times with short time positive auto-correlation in velocity. The persistence length of their trajectories is given by $||U_0|| / D_r$ where $D_r$ is the rotational diffusion coefficient.

\subsection{Rotations}

\begin{figure}[h]
\centerline{\includegraphics[width=8 cm]{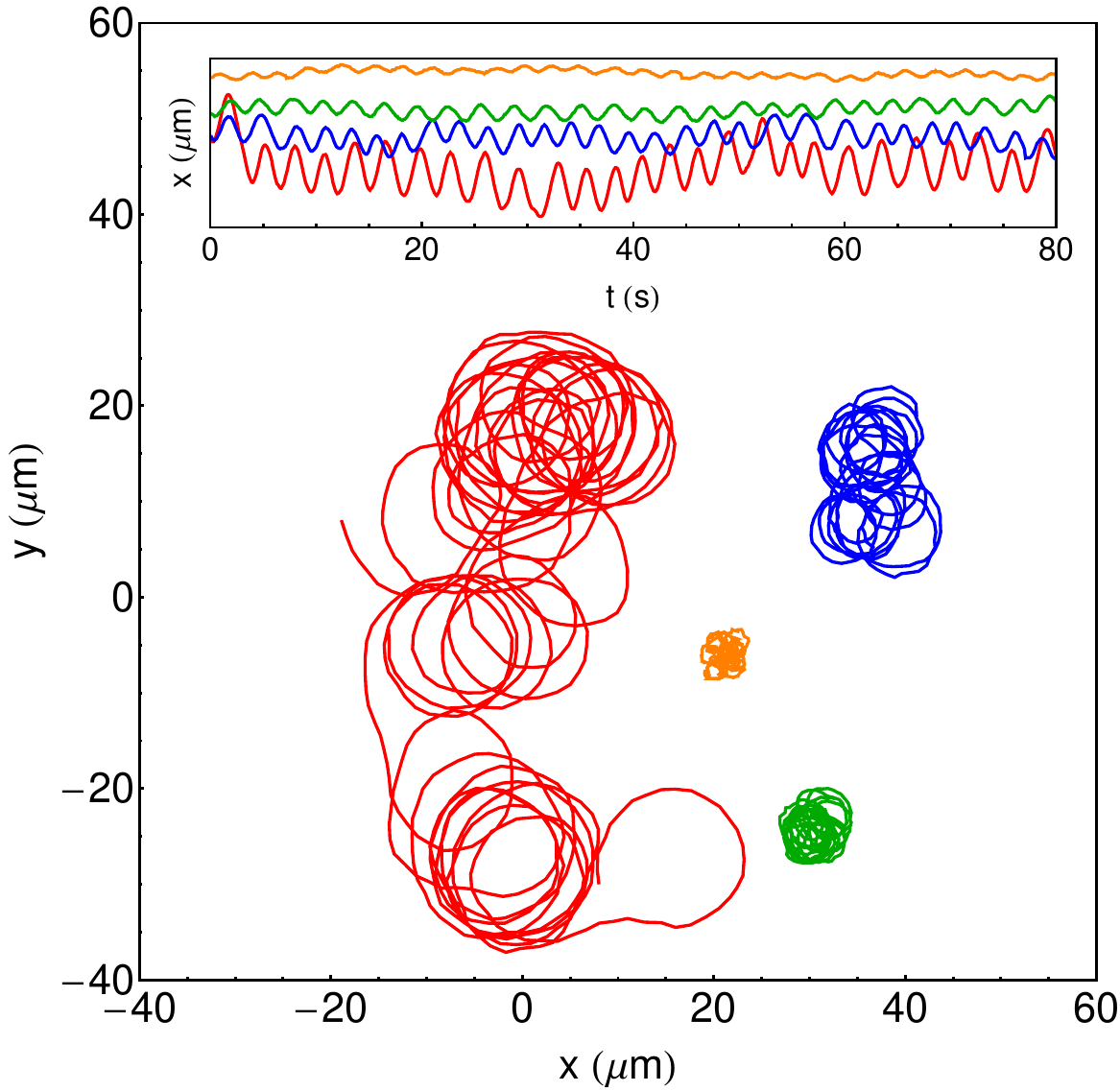}}
\caption{\label{rotations}Rotations of a Janus particle for $E_0 = 6 \mbox{ V}$ and $\omega = 3$ (red), $6$ (blue), $10$ (green) and $15\mbox{ kHz}$ (orange). The inset corresponds to the time evolution of the projection of these trajectories on the x axis.}
\end{figure}

Instead of ABM, we found that some particles would rather exhibit noisy rotations, moving in circular trajectories at a constant frequency $\Omega$ (see Fig.~\ref{rotations}). The direction of rotation (clockwise or counter-clockwise) depends on the particle and never changes once the rotations have been initiated: a particle turning clockwise will keep turning clockwise as long as the electric field is on. Therefore, each particle has its own rotation axis. However, it also depends on the orientation of the particle at $t = 0$, when $E$ is applied: a particle turning clockwise might turn counter-clockwise if we turn off the electric field, wait a few seconds and then turn it on again. This might be caused by turning the particle upside down along the z-axis. A given particle can often switch between ABM and rotations: typically, it will exhibit ABM at high frequencies $\omega$ and small amplitudes $E_0$ and rotations at low $\omega$ and high $E_0$. However, it should be noted that the transition values $(\omega^s, E_0^s)$ for which the switching occurs are specific to each particle.
How can we explain these rotations? It is clear that the Janus particles are not only subjected to a self-propelling force $\boldsymbol{F}$ but also to a torque $\boldsymbol{M}$ inducing the rotations. If we look at the particles at very high magnification using an electron microscope, the frontier between the metal side and the polystyrene side does not appear perfectly straight (not shown here, see Fig. 1 of reference \cite{Merkt06} for example). The two quadrants of metal coated hemisphere separated by the $(xz)$ plane can then be covered by different quantities of metal as it is shown on the inset of Fig.~\ref{scheme_Janus}. Therefore, we can assume that each Janus particle has not only breaking front-back 
symmetry but also breaking chiral symmetry so that the ICEO flow will also be unbalanced with respect to $(xz)$ mirror plane.
\begin{figure}
\begin{tabular}{cc}
\includegraphics[width=4.4 cm]{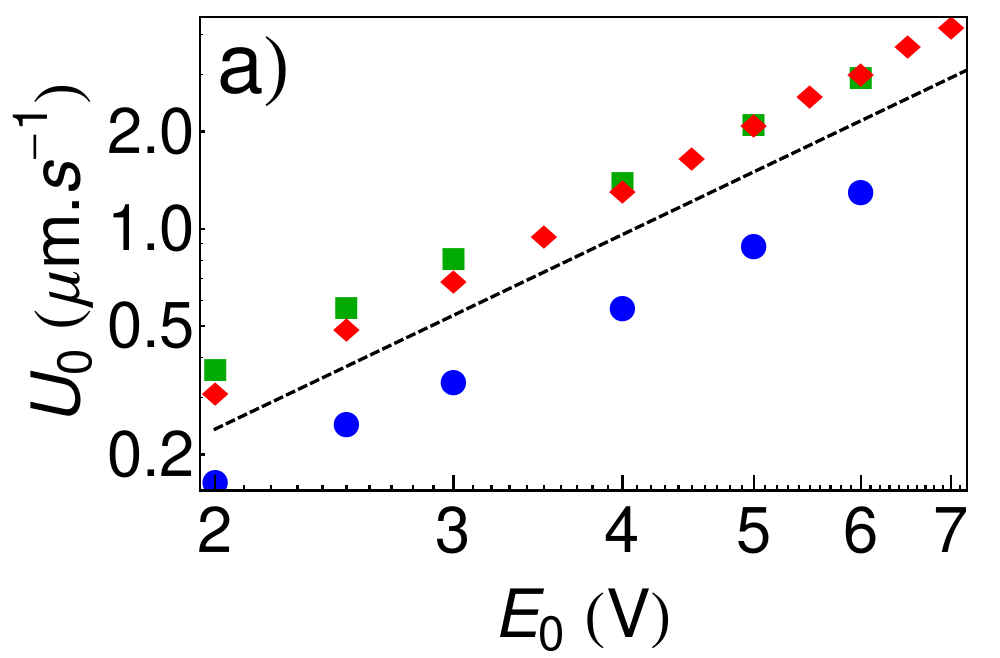} &
\hspace{-0.5cm}
\includegraphics[width=4.4 cm]{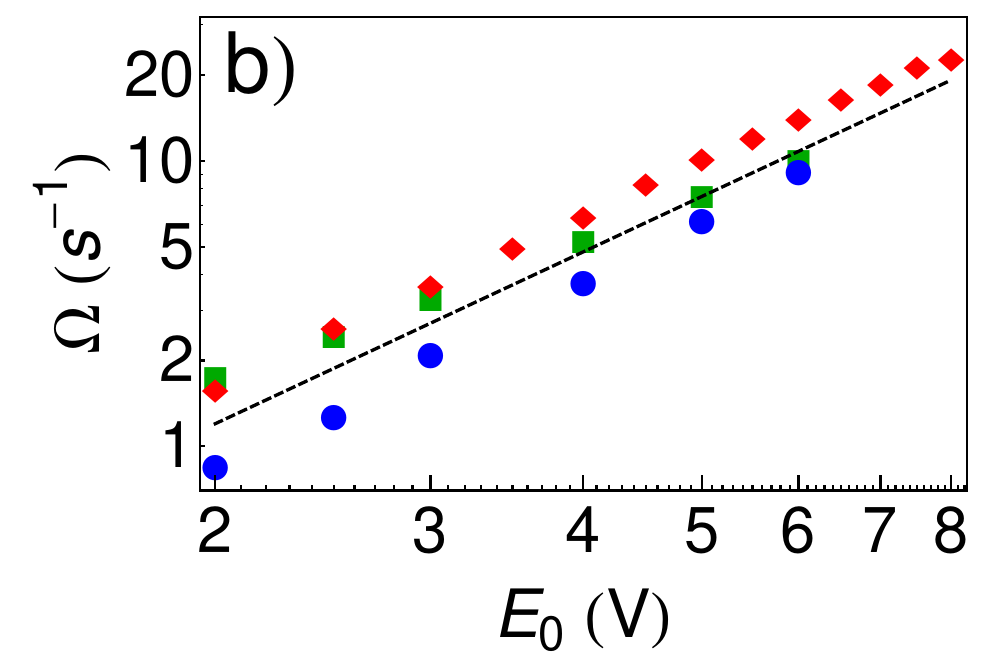} \\
\includegraphics[width=4.4 cm]{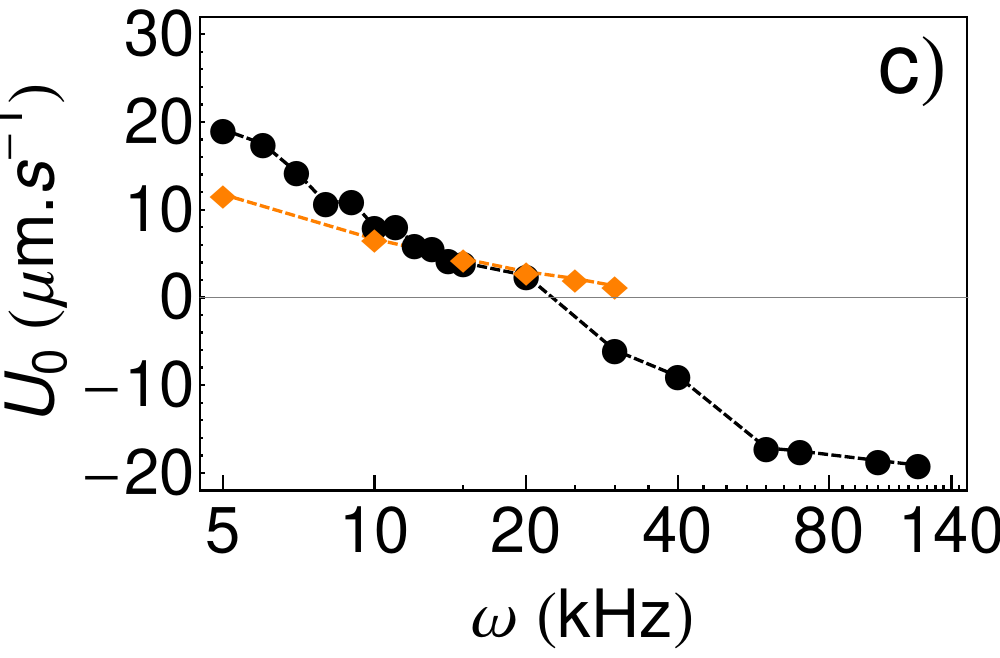} &
\hspace{-0.5cm}
\includegraphics[width=4.4 cm]{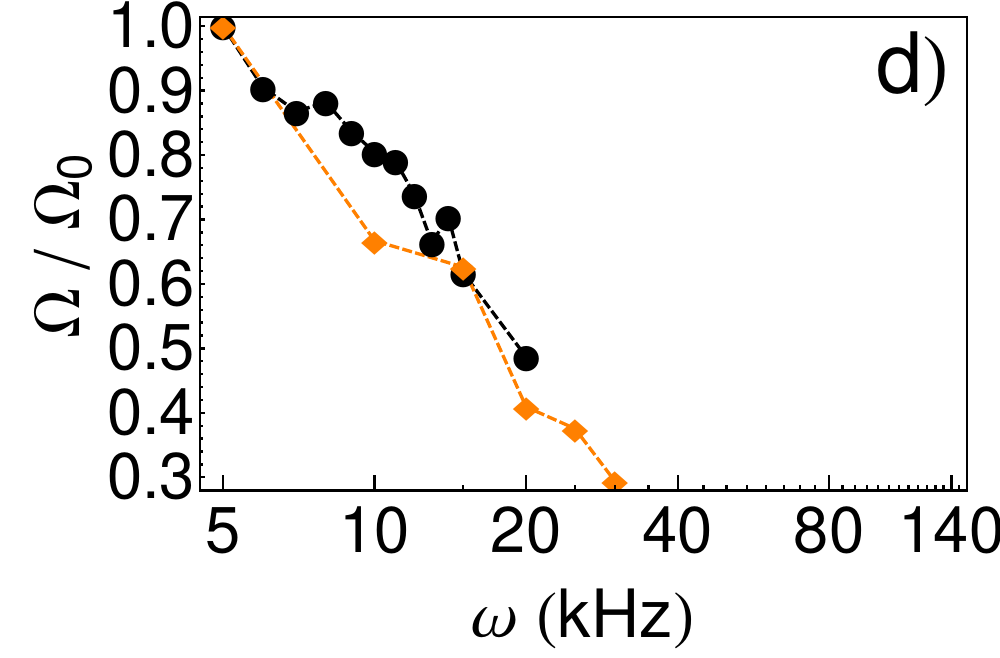}
\end{tabular}
\centering{\includegraphics[width=4.4 cm]{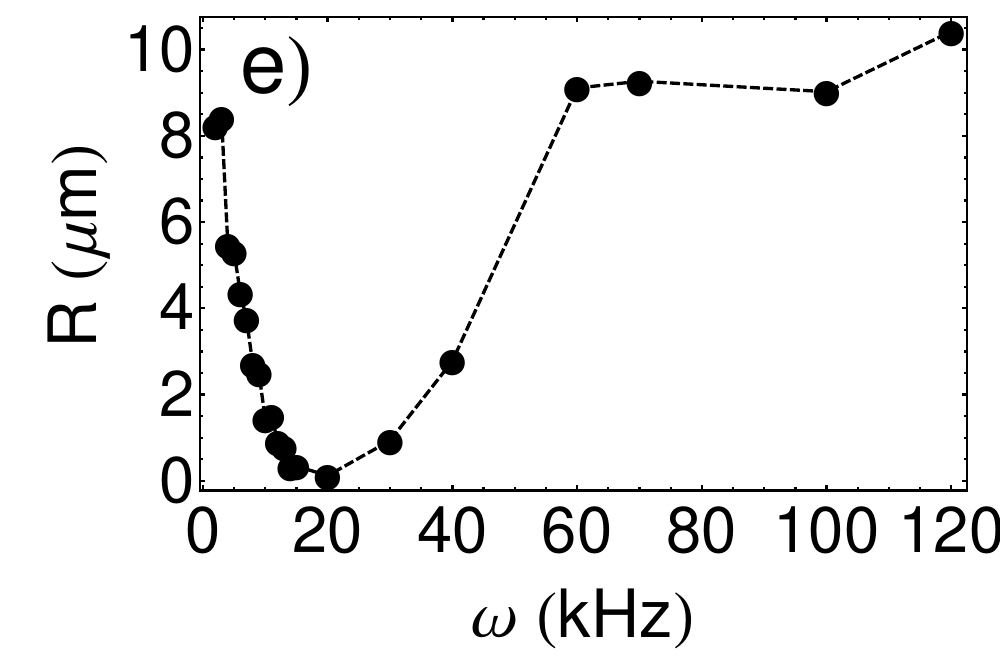}}
\caption{\label{U0_Omega}Examples of the evolution of the mean velocity $U_0$, the rotational frequency $\Omega$ and the radius of rotation $R$ with respect to the two control parameters $E_0$ and $\omega$ for several particles. The dashed lines on Figs. a) and b) are of slope $2$, highlighting that $U_0$ and $\Omega \propto E_0^2$. When the data points of Fig. c) cross the red line, the direction of propagation of the particles is reversed. The symbols of different colors correspond to different particles. a) and b) $\omega = 10 \mbox{ kHz}$, c), d) and e) $E_0 = 6$. In Fig. d), $\Omega$ has been nondimensionalized by its value at 5 kHz $\Omega_0$ with $\Omega_0 = 0.46 \mbox{ s}^{-1}$ (orange diamonds) and $2.62 \mbox{ s}^{-1}$ (black disks).a) and b) are log-log plots, c) and d) semi-log ones and e) a linear one.}
\end{figure}
\noindent The strength of the resulting torque depends on how asymmetric a given particle is with respect to $(xz)$ and its sign on the initial orientation of the particle. Note that the possibility of chiral Janus particle produced by coating imperfection has been mentioned in the literature\cite{Cordes2015}.\\

The experimental values of $\boldsymbol{U_0}$ and $\Omega$ can be extracted from the trajectories of the particles. The motion of a Janus particle can be accurately described by the following system of coupled Langevin equations:
\begin{equation}
\label{Langevin}
\left\{
\begin{array}{ll}
\dot{\boldsymbol{r}} =  \boldsymbol{U_0} + D_t \boldsymbol{\xi}_t \\
\dot{\phi} = \Omega +  D_r \xi_r,
\end{array}
\right.
\end{equation}
with $D_t$ and $D_r$ the translational and rotational diffusion coefficients, $\boldsymbol{\xi}_t$ and $\xi_r$ the translational and rotational noises and $\phi$ the angle of the orientation unit vector $\boldsymbol{\hat{u}}$ such that $\boldsymbol{U_0} = U_0 \boldsymbol{\hat{u}} = U_0 (\cos{\phi} \, \boldsymbol{\hat{e}_x} + \sin{\phi} \, \boldsymbol{\hat{e}_y})$. This system can be solved exactly to get the analytical expression of the velocity auto-correlation function (see the appendix):
\begin{equation}
\label{vcorr}
\langle \boldsymbol{v}(t) \cdot \boldsymbol{v}(t+\tau)\rangle = U_0^2 \,\exp(-D_r \tau) \, \cos(\Omega \, \tau).
\end{equation}
The average velocity $U_0$ and rotational frequency $\Omega$ of the particles are directly proportional to the force and torque respectively:
\begin{equation}
\left\{
\begin{array}{ll}
U_0 &= F / (m \gamma)\\
\Omega &=  4 M / (m \gamma d^2),
\end{array}
\right.
\end{equation}
with $m$ the mass of the particle, $\gamma$ the damping constant and $d$ the diameter of the particle. Therefore, if $\boldsymbol{F}$ and $\boldsymbol{M}$ have the same origin, we should have $\Omega \propto E_0^2$ and $U_0 \propto E_0^2$, according to the ICEP theory. Eq.~[\ref{vcorr}] can be used to fit the experimental curves of the auto-correlation function thus extracting the values of $\Omega$ and $D_r$ (see Fig.~\ref{correlations}). The evolution of $U_0$ and $\Omega$ with respect to $E_0$ are shown on Figs.~\ref{U0_Omega} a) and b). Both of them are proportional to $E_0^2$, confirming that $F$ and $M$ have indeed the same origin.\\
Figs.~\ref{U0_Omega} c) and d) show that the evolution of $U_0$ and $\Omega$ with respect to $\omega$ is more particle-dependent. On average, $U_0$ and $\Omega$ decrease with respect to this parameter. Note that $U_0$ decreases more quickly than $\Omega$ with respect to $\omega$. The type of motion of the particle can be characterized by its rotation radius $R(\omega)=||U_0||/\Omega$ ($R$ does not depend on $E_0$ since both $U_0$ and $\Omega$ are proportional to $E_0^2$). When $R(\omega)$ is about the size of the particle $d$, the motion is considered rotational. On the other hand, if $R(\omega) \gg d$, the motion is dominated by self-propelling force and equivalent to an ABM. The evolution of $R(w)$ is non-monotonic as shown in Fig.~\ref{U0_Omega} e): the radius of rotation first decreases when $\omega$ increases because $U_0$ decreases more rapidly than $\Omega$ in the SP region. When switching to the ISP region, $R(w)$ increases again and for high frequencies (typically $\ge 150 \mbox{ kHz}$), the motion of the particles becomes similar to an ABM so that it becomes difficult to measure $R$ or $\Omega$.

\section{Feedback control of a Janus particle}

\begin{figure*}
\begin{tabular}{cc}
\centerline{\includegraphics[width=16 cm]{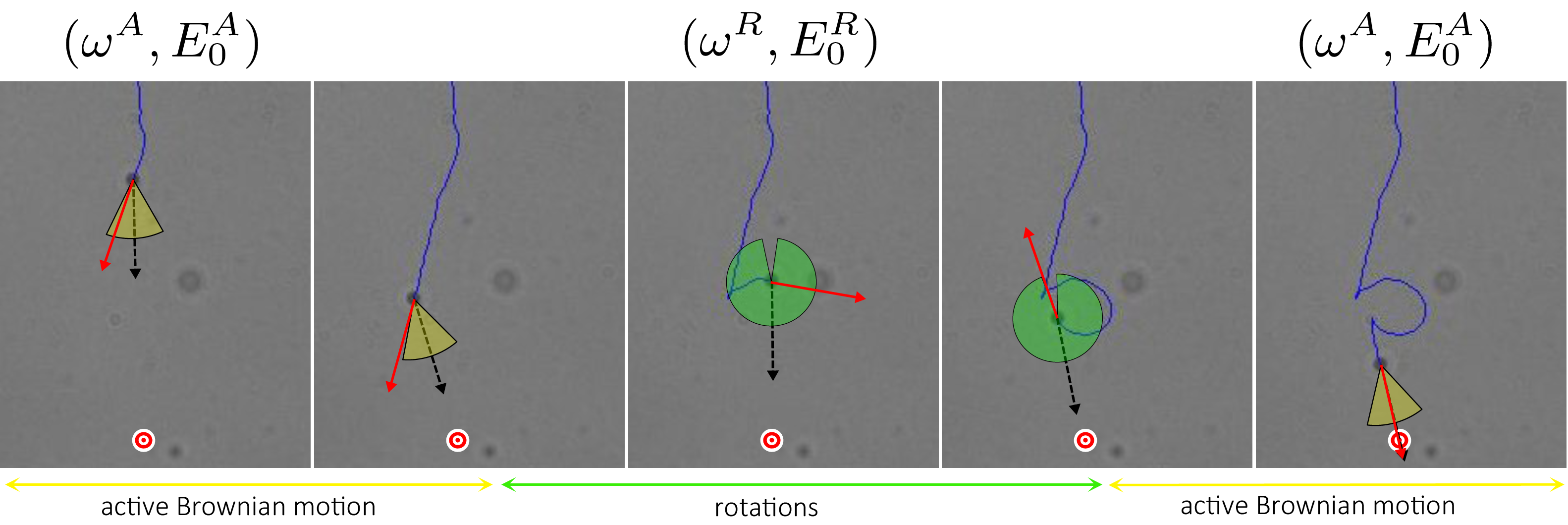}}
\end{tabular}
\caption{\label{feedback}Schematic picture representing the algorithm of feedback manipulation. The concentric red and white disks are the target. At each frame, we plot the trajectory of the particle (in blue) and its instantaneous velocity vector $\tilde{\boldsymbol{v}}(t)$ (red arrow). We then compare it to the vector pointing at the target $\boldsymbol{v}_{T}(t)$ (black dotted arrow), given the tolerance angles we have chosen: $\alpha = 0.7 \approx 40^{\circ}$ (yellow circular sector) and $\alpha_{R} = 2.97 \approx 170^{\circ}$ (green circular sector). If $\tilde{\boldsymbol{v}}(t)$ is included in the yellow circular sector, the particle exhibits ABM and its direction of propagation is more or less correct. If $\tilde{\boldsymbol{v}}(t)$ is included in the green circular sector, the particle rotates but its orientation still needs to be reoriented. Thus, the control parameters are only switched when $\tilde{\boldsymbol{v}}(t)$ gets out of the two circular sectors. Note that when this happens, the direction of propagation of the particle is reversed (which is the reason why we chose this value for $\alpha_{R}$).}
\end{figure*}

The two distinct behaviors we have described so far - switching between ABM and rotations - remind us of the well studied motion displayed by some bacteria: the ``run-and-tumbling'' chemotaxis. Inspired by this kind of motion, we present and demonstrate a novel method to transport an individual Janus particle to a given position in the 2D space. The key of this method is to control the direction of propagation of a given particle by switching between ABM and rotations by shifting the AC frequency of the applied electric field $\omega$ at the right moment (see Fig.~\ref{feedback}): when the orientation vector of the particle $\boldsymbol{\hat{u}}$ is directed to the target, $\omega$ is set to a high value so that ABM is induced. As soon as the particle is misdirected due to the thermal noise, $\omega$ is set to low values and the particle starts rotating. Its orientation vector then evolves continuously until it points once again to the target, at which moment the particle is set back to the ABM state. 

Based on this simple idea, we developed a computer program that tracks a particle in real time and automatically applies the right control parameter to the system (for detailed implementation, see Materials and Methods). Suppose we wish to transport a particle to a certain target located at $\boldsymbol{R}_{T}$ in the two-dimensional plane. At each frame captured by the camera, our program does the following operations:
\begin{enumerate}
\item get the current position $\boldsymbol{r}(t)$\\
\item compute its smoothed instantaneous velocity 
\begin{equation*}
\tilde{\boldsymbol{v}}(t)= \sum_{n=0}^{N_f} \left[ \boldsymbol{r}( t-n\Delta t ) - \boldsymbol{r}( t-(n+1)\Delta t) \right]/(N_f+1).
\end{equation*}
$N_f$ is the number of frames used to smooth the instantaneous velocity. We assume $\tilde{\boldsymbol{v}}(t)$ is roughly parallel to the orientation vector of the particle $\boldsymbol{\hat{u}}(t)$\\
\item compute the vector pointing at the target $\boldsymbol{v}_{{T}}(t) = \boldsymbol{R}_{{T}} - \boldsymbol{r} (t)$\\ and calculate the angle between the two vectors
\begin{equation*}
\theta(t) =\arccos{\left[ \frac{\tilde{\boldsymbol{v}}(t) \cdot \boldsymbol{v}_{T}}{||\tilde{v}(t)|| \, ||v_{T}||} \right]}.
\end{equation*}
\begin{enumerate}
\item[4.a] if the particle is in ABM state, compare with the ABM tolerance angle $\alpha$. If $\theta \leq \alpha$, the particle stays in ABM state. Otherwise, the particle is switched to rotation state by setting the control parameter to $(\omega^{R}, E_0^{R})$.
\item[4.b] if the particle is in rotation state, compare with the rotation tolerance angle $\alpha_R$. If $\theta \leq \alpha_R$, the particle stays in rotation state. Otherwise, the particle is switched to ABM state by setting the control parameter to $(\omega^{A}, E_0^{A})$.
\end{enumerate}
\item[5.] repeat from 1.
\end{enumerate}
As we have seen in the previous section, the values of $(\omega^{A}, E_0^{A})$ and $(\omega^{R}, E_0^{R})$ are specific to each particle. However just like biological chemotaxis is robust to variability to gene expression or fluctuating environmental conditions, this algorithm can be used to control most of the particles regardless of their variability. To achieve a most efficient transportation of particles, one needs to choose the parameters $(\omega^{A}, E_0^{A})$ and $(\omega^{R}, E_0^{R})$ so that they maximize the persistence length and minimize the rotation radius. In other words, we want $R \gg d$ and $U_0 / D_r \gg d$ for $(\omega^{A}, E_0^{A})$ and $R \ll d$ for $(\omega^{R}, E_0^{R})$. Considering the characterization of the motion described in the previous part, we will use high amplitudes for $E_0^{R}$, high frequencies for $\omega^{A}$ and small frequencies $\omega^{R}$. The precise values of these parameters have to be determined manually for each particle before initiating the feedback control and typically, $E_0^R = E_0^A \approx 7 \mbox{ V}$, $\omega^{A} \approx 300 \mbox{ kHz}$ well above the cross over frequency and $\omega^{R} \approx 5 \mbox{ kHz}$ low enough to have small $R$. It is important to note that because we switch from low to high frequencies, we go from the SP to the ISP region and the direction of the motion is reversed. Regarding the tolerance angles, we typically used $\alpha = 0.7 \approx 40^{\circ}$ and $\alpha_{R} = 2.97 \approx 170^{\circ}$. In an ideal case, $\alpha_R$ should be equal to $\pi$. However, because $\tilde{\boldsymbol{v}}(t)$ is averaged over $N_f$ frames, there is always a small delay between $\tilde{\boldsymbol{v}}(t)$ and the actual orientation of the particle $\boldsymbol{\hat{u}}(t)$. To take this delay into account, we use a slightly smaller value for $\alpha_R$. Using a smoothed instant velocity is important to decrease the sensitivity of the algorithm to translational noise.
\begin{figure}
\begin{tabular}{cc}
\centerline{\includegraphics[width=7 cm]{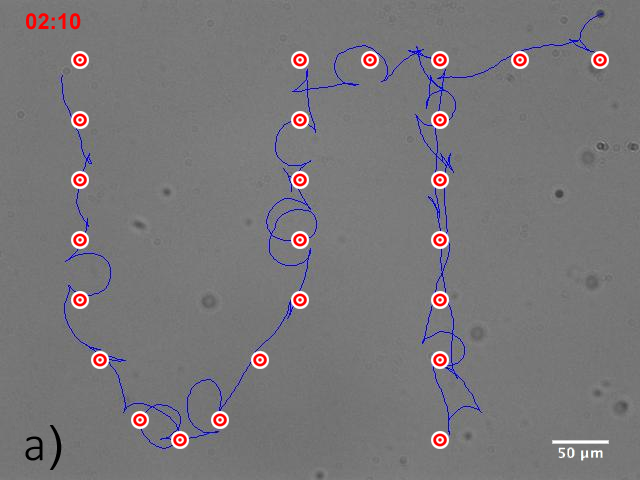}}\\
\centerline{\includegraphics[width=7 cm]{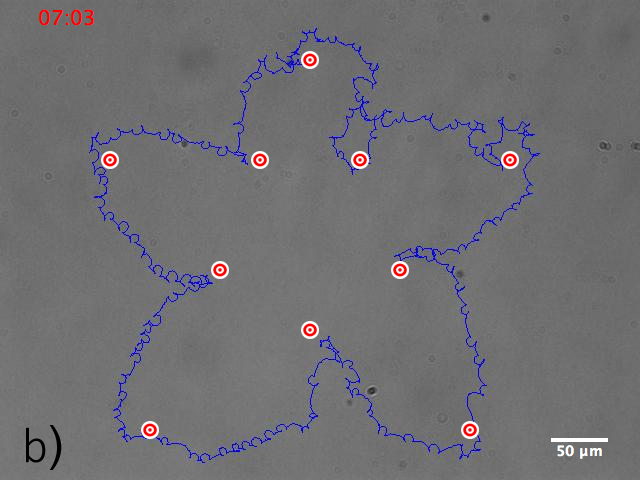}}
\end{tabular}
\caption{\label{sakura}Examples of ``Microtag'': a) the letters ``UT''. b) a cherry blossom flower. Both were obtained using the feedback-control program and a list of target locations which appear as concentric red and white disks. The trajectory of the particle appears in blue.}
\end{figure}
With this algorithm, we were successfully able to direct the motion of our Janus particles (see movies in the supplementary material). The ``active'' reorientation process is particularly efficient: its typical time scale $\Omega^{-1}$ is less than a second, which is much faster than a ``passive'' reorientation by rotational diffusion, determined by $D_r^{-1} = 8\pi \eta d^3 / k_B T \approx 20 \mbox{ s}$ for a spherical particle of radius $d = 1.5 \, \mu \mbox{m}$. Moreover, we do not have to be able to see the orientation of the particles to reorient them as we can deduce $\hat{\boldsymbol{u}}(t)$ from their instantaneous velocity. The present method has an advantage if the orientation of a Janus particle is not easily accessible due to smallness of its size. The method enables us not only sending a particles to a given target, but also designing a trajectory of particle to a certain extent by giving a list of target coordinates sequentially. We were thus able to realize ``microtags'' as can be seen in Fig.~\ref{sakura}. The control of the trajectory is of course not perfect as the particle makes a turn with a small radius as visible on the trajectory having small loops in Fig.~\ref{sakura}.

\section{Optimal feedback strategy}

\begin{figure}
\begin{tabular}{cc}
\includegraphics[width=8 cm]{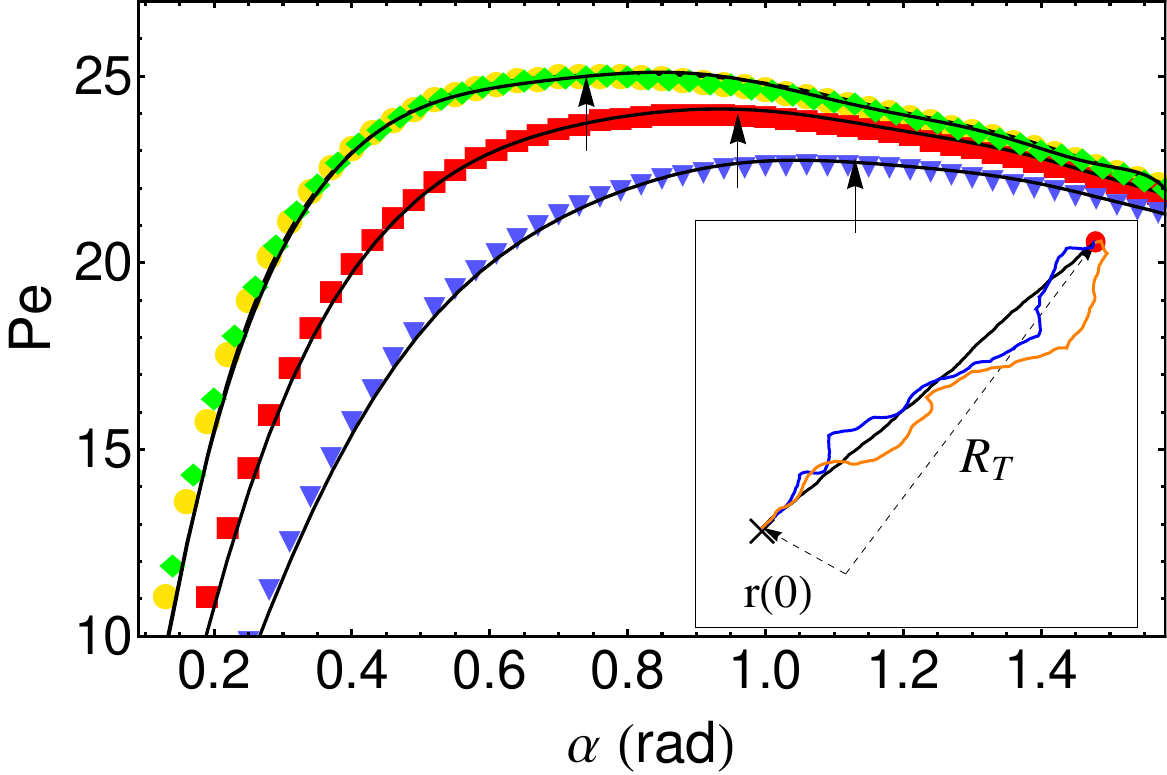}
\end{tabular}
\caption{\label{optimum}Evolution of the P\'eclet number with respect to $\alpha$ for a target at distance $r = 98\sqrt{2} \mbox{ } \mu \mbox{m}$ and $U_0 =0.2 \mbox{ } \mu \mbox{m.s}^{-1}$. Yellow) $D_r = 0.2 \mbox{ s}^{-1}$ and $\Omega = 40 \mbox{ s}^{-1}$. Red) $D_r = 0.2 \mbox{ s}^{-1}$ and $\Omega = 20 \mbox{ s}^{-1}$. Blue) $D_r = 0.2 \mbox{ s}^{-1}$ and $\Omega = 10 \mbox{ s}^{-1}$. Green) $D_r = 0.1 \mbox{ s}^{-1}$ and $\Omega = 20 \mbox{ s}^{-1}$. $\alpha_R = 0$ for all the results shown here. The black curves correspond to the theoretical predictions given by Eq.~[\ref{Peclet2}] and the arrows indicate the maximum P\'eclet number. Inset: examples of numerical trajectories of a particle under feedback control for several values of tolerance angle: $\alpha = 0.3$ (black), $1.1$ (blue) and $1.5$ (orange). The red disk corresponds to the position of the target. b)}
\end{figure}

For the present algorithm, the efficiency of the feedback depends on the tolerance angle $\alpha$. In principle, the feedback can be optimized by choosing the most appropriate value of $\alpha$. One way to quantify its efficiency is to measure the P\'eclet number, defined as the ratio between the time needed to diffuse a given distance and the time required to swim the same distance. To calculate it, we need to use the average velocity of a particle under feedback control $\langle v \rangle$ given by : 
\begin{equation}
\label{v_mean}
\langle v \rangle = \frac{\langle \Delta x_{A} \rangle}{\langle \Delta t_{A} \rangle + \langle \Delta t_{R} \rangle},
\end{equation}
where $\langle \Delta t_{A} \rangle$ and $\langle \Delta t_{R} \rangle$ correspond to the average durations of one cycle of ABM and of rotations respectively and $\langle \Delta x_{A} \rangle$ is the average displacement during one cycle of ABM projected on the axis pointing to the target. The P\'eclet number is thus equal to
\begin{eqnarray}
\label{Peclet}
\mbox{Pe} &=& \frac{r_0^2}{D_t} \frac{\langle v \rangle}{r_0} = \frac{r_0 \langle \Delta x_A \rangle}{D_t(\langle \Delta t_A \rangle +\langle \Delta t_R \rangle)}\\
&=& \frac{r_0 U_0}{D_t} \frac{\langle v \rangle}{U_0} = \mbox{Pe}_L \, \langle \cos \theta \rangle ,
\label{Peclet_cos}
\end{eqnarray}
with $r_0 = ||\boldsymbol{R}_{T}-\boldsymbol{r}(0)||$ the initial distance between the particle and the target and $\mbox{Pe}_L = r_0 U_0/D_t$ the mass P\'eclet number. Eq.~[\ref{Peclet_cos}] shows that $\mbox{Pe}$ is directly proportional to $\langle \cos \theta \rangle$, the chemotaxis index generally used to measure the accuracy of chemotaxis\cite{Endres08}. To understand how the efficiency of the feedback control depends on $\alpha$, we ran numerical simulations by integrating the system of Eqs.~[\ref{Langevin}] using a standard Gillespie algorithm. We made several simplifying assumptions: we neglected the influence of the translational noise, did not consider the reversion of self-propulsion and assumed that the radius of rotation is so small that the particle does not actually move translationally in the rotation state. When the control parameters are $(\omega^{A}, E_0^{A})$, the motion of the particle under feedback control is thus described by
\begin{equation}
\label{sys_Langevin_RW}
\left\{
\begin{array}{ll}
\dot{\boldsymbol{r}} = \boldsymbol{U_0}\\
\dot{\phi} = D_r \xi_r.
\end{array}
\right.
\end{equation}
On the other hand, when the control parameters are $(\omega^{R}, E_0^{R})$ the equations of motion become
\begin{equation}
\label{sys_Langevin_R}
\left\{
\begin{array}{ll}
\dot{\boldsymbol{r}} = 0 \\
\dot{\phi} =  \Omega + D_r \xi_r.
\end{array}
\right.
\end{equation}
Integrating these two systems of equations, we obtained $200$ trajectories of particles for various tolerance angles and extracted the values of $\langle \Delta x_{A} \rangle$, $\langle \Delta t_{A} \rangle$ and $\langle \Delta t_{R} \rangle$. Typical trajectories of these simulations are shown in the inset of Fig.~\ref{optimum}. At first glance, small tolerance angles seem to be optimum as they correspond to the shortest path to the target. However, Fig.~\ref{optimum} shows that small values of $\alpha$ actually correspond to small P\'eclet numbers. Indeed, there is a balance to be found between taking the shortest path to the target and wasting too much time in reorientation process. Note that the optimum angle depends on the value of the angular frequency $\Omega$ and on the value of $D_r$.

Obtaining a theoretical estimate of the P\'eclet number with respect to the tolerance angle is quite cumbersome thus we use a crude approximation: we assume that the particle is heading right to the target in the ``run'' state. In other words, we neglect the curvature of the trajectory, which is reasonable for small tolerance angles or if the particle is very far away from the target. This is the case in our simulations as the persistence length $U_0 / D_r$ is much smaller than the initial distance to the target. Therefore, the angle between the particle and the target will only be modified by the rotational diffusion (which means that $\theta(t) = \phi(t)$). If a particle initially has the right orientation ($\phi(t=0) = \phi_0 = 0$), it will thus need to be reoriented again after an average time given by:
\begin{equation}
\label{delta_t_RW}
\langle \Delta t_{A} \rangle = \frac{\alpha^2}{2 D_r}.
\end{equation}
A simple integration of the first equation of~[\ref{sys_Langevin_RW}] gives us
\begin{equation}
\begin{array}{ll}
\Delta x(t) & = U_0 \int^t_0 \cos{\phi(t')} dt'.
\end{array}
\end{equation}
The average displacement $\langle \Delta x_{A} \rangle$ is thus given by
\begin{equation}
\label{deltax}
 \langle \Delta x_{A} \rangle = U_0 \int^{\infty}_0 \mbox{FPTD}(t, \alpha) \int^{\alpha}_{-\alpha} \int^{t}_0 P(\phi,t')\cos{\phi(t')} \, dt' d\phi  \, dt,
\end{equation}
where we have averaged using the first-passage time distribution $\mbox{FPTD}(t, \alpha)$ and the probability density function of $\phi$ $P(\phi,t')$. $P(\phi,t')$ is the normalized PDF of a one-dimensional variable only subjected to thermal fluctuations, with absorbing boundary conditions at $\phi = \pm \alpha$. Therefore, we must have at all times
\begin{eqnarray}
\label{condition}
 P(|\phi| \ge \alpha,t) &= 0 \\
\int_{-\alpha}^{\alpha}  P(\phi,t) \, d\phi & = 1.
\label{condition2}
\end{eqnarray}
Using the mirror image method\cite{Zauderer11}, it is easy to show that $P(\phi,t)$ is equal to
\begin{equation}
\label{phi_AB}
\begin{split}
 P(\phi,t) =  \frac{1}{C(\alpha,t)} \frac{1}{\sqrt{4 \pi D_r t}} \sum_{n=-\infty}^{\infty} \left[ \exp \left( -\frac{(\phi + 4 n \alpha)^2}{4 D_r t} \right) \right.\\
 \left. - \exp \left( \frac{(\phi + (4 n-2) \alpha)^2}{4 D_r t} \right) \right],
\end{split}
\end{equation}
where $C(\alpha,t)$ is a normalization function required to satisfy Eq.~[\ref{condition2}] and given by 
\begin{equation}
\label{norm}
\begin{split}
C(\alpha,t) = \frac{1}{2} \sum_{n=-\infty}^{\infty} \left[ \erf \left( -\frac{(- 3 + 4 n) \alpha}{2 \sqrt{D_r t}} \right) \right. \hspace{2cm}\\
 \left. - 2 \, \erf \left( -\frac{(- 1 + 4 n) \alpha}{2 \sqrt{D_r t}} \right) + \erf \left( -\frac{(1 + 4 n) \alpha}{2 \sqrt{D_r t}} \right) \right] .
\end{split}
\end{equation}
As for the first-passage time distribution $\mbox{FPTD}(t, \alpha)$, it is given by the following equation for these boundary conditions\cite{Darling53}
\begin{equation}
\label{fptd}
\begin{split}
\mbox{FPTD}(t, \alpha) =  \frac{\pi D_r}{\alpha^2} \sum_{n=0}^{\infty} (-1)^n  (2n+1)  \hspace{3.4cm} \\
\cos \left( \frac{(2n+1)\pi}{2 \alpha} \phi_0\right)\exp \left( - \left[ \frac{(2n+1) \pi}{2 \alpha} \right]^2 t \, D_r \right).
\end{split}
\end{equation}
Injecting Eqs.~[\ref{fptd}] and~[\ref{phi_AB}] into~[\ref{deltax}] and using the change of variable $D_r \, t \to T$, we finally obtain   
\begin{equation}
\begin{split}
 \langle \Delta x_{A} \rangle = \frac{U_0}{D_r} I(\alpha) \hspace{0.5cm} \mbox{with} \hspace{8cm}\\
I(\alpha) = \int^{\infty}_0 \int^{\alpha}_{-\alpha} \int^{T}_0 \mbox{FPTD}\left(\frac{T}{D_r}, \alpha \right) \, P\left(\phi,\frac{T'}{D_r}\right) \hspace{4.2cm}\\ 
\cos{\phi} \, d\phi \, dT' \, dT. \hspace{6.5cm}
\end{split}
\end{equation}
Note that the integral $I(\alpha)$ only depends on the value of the tolerance angle. We could not find an analytical expression for it but it can be evaluated numerically.\\

Let us now focus on Eqs.~[\ref{sys_Langevin_R}]. Integrating the second equation gives us the evolution of the orientation angle with respect to the time
\begin{equation}
\langle \phi(t) \rangle = \phi_0 + \Omega t.
\end{equation}
To correct the orientation of the particle, $\phi(t)$ needs to be changed either by $2\pi - \alpha_{R} - \alpha$ or $\alpha - \alpha_{R}$, depending on the sign of $\Omega$. The average reorientation time is thus given by 
\begin{equation}
\langle \Delta t_{R} \rangle = \frac{\pi -\alpha_{R}}{\Omega}.
\end{equation}

\begin{figure}
\begin{tabular}{cc}
\includegraphics[width=8 cm]{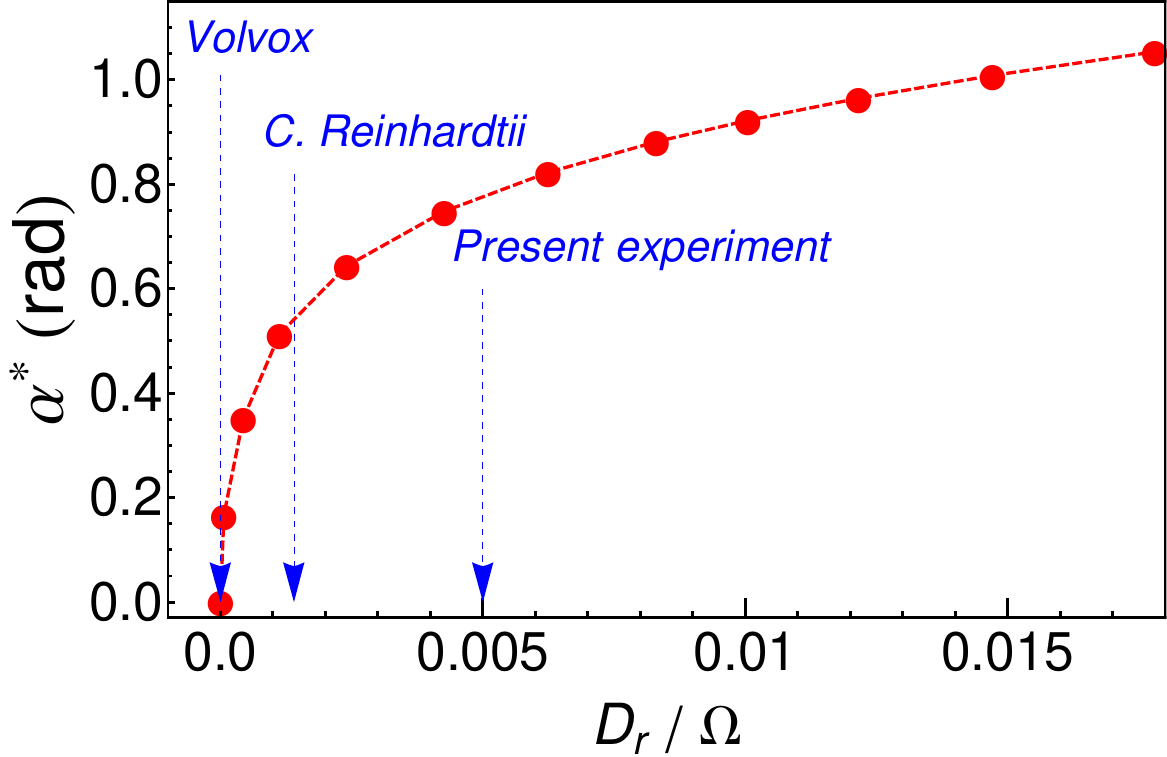}
\end{tabular}
\caption{\label{diagram}Evolution of the optimum tolerance angle $\alpha^*$ with respect to $D_r / \Omega$ for $\alpha_R = 0$. The blue dashed arrows show typical values of $D_r / \Omega$ for our Janus particles, \textit{Chlamydomonas reinhardtii} and \textit{Volvox}. \textit{Escherichia coli} is not shown here as it corresponds to a greater value ($D_r / \Omega = 0.024$) .}
\end{figure}

Using all the previous results, the equation for the P\'eclet number becomes
\begin{equation}
\label{Peclet2}
\mbox{Pe} = \frac{U_0 \, r_0}{D_r \, D_t} \left(\frac{\alpha^2}{2 D_r} + \frac{\pi -\alpha_{R}}{\Omega}\right)^{-1} I(\alpha) .
\end{equation}
We have compared it to the results of simulations. As we can see on Fig.~\ref{optimum}, the agreement is excellent. We can thus determine the optimum value of tolerance angle $\alpha^*$ corresponding to the maximum P\'eclet number for a given set of control parameters. According to Eq.~[\ref{Peclet2}], $\mbox{Pe}$ is linearly proportional to $U_0$ and depends on $\alpha$ and on the ratio $D_r / \Omega$ (see Fig.~\ref{optimum}). Therefore, $\alpha^*$ should be a function of this sole parameter $D_r / \Omega$,  i.e. signal-to-noise ratio. This is in good agreement with analytical results obtained for the chemotaxis of the bacteria \textit{Escherichia coli} that showed that the optimum angle for their run-and-tumbling motion only depends on the ratio $D_r \tau_{tumble}$, where $\tau_{tumble}$ is the average time required to reorient the bacteria\cite{Strong98}. In our case, this time is clearly determined by the strength of the torque $\Omega$ so that the efficiency depends instead on $D_r / \Omega$. Fig.~\ref{diagram} highlights that fast reorientations or weak noises (low $D_r / \Omega$) are associated with small optimum angles $\alpha^*$. For very slow reorientations ($D_r/\Omega \to \infty$), $\alpha^*$ saturates at $\pi/2$. Note that for reasonable values of the parameters ($D_r = 0.05 \mbox{ s}^{-1}$, $U_0=2 \, \mu \mbox{m.s}^{-1}$ and $\Omega = 10 \mbox{ s}^{-1}$), $\alpha^* \approx 0.78 \approx 45^{\circ}$ which is quite close to what we had determined empirically in our experiments.\\
In biology and biophysics, strategies of run-and-tumble and steering motion have been discussed for different sizes of swimming organisms such as \textit{Escherichia coli}\cite{Strong98}, \textit{Chlamydomonas reinhardtii}\cite{Polin09}, and \textit{Volvox}\cite{Drescher10}. Therefore, it is interesting to compare them in regard to the signal to noise ratio depicted in Fig.~\ref{diagram}. In the case of \textit{Escherichia coli} ($r = 1 \, \mu \mbox{m}$, $D_r = 0.16 \mbox{ s}^{-1}$, $\tau_{tumble}=0.15 \mbox{ s}$), $\alpha^* \sim 1.1 \mbox{ rad} \sim 63^{\circ}$ can be deduced if we assume that $\Omega = (\tau_{tumble})^{-1}$. However, \textit{Escherichia coli} does not have directional sensors nor steering control and its flagella can only reverse the direction of rotation. Thus, it takes a strategy of random reorientation. \textit{Chlamydomonas reinhardtii} ($r = 5 \, \mu \mbox{m}$, $D_r = 0.0014 \mbox{ s}^{-1}$, $\Omega=1 \mbox{ s}^{-1}$, $D_r / \Omega = 0.0014$) is 2 times larger than our Janus particle, and 5 times larger than \textit{Escherichia coli} thus 100 times less sensitive to rotational diffusion. It has a photoreceptor and could have steering strategy, but in fact it also takes run-and-tumble strategy. It was discovered that \textit{Chlamydomonas reinhardtii} can switch synchronized and non-synchronized beating between two flagella for making straight and tumbling motion respectively\cite{Polin09}. This was explained as a trade-off between the resource exploration and the avoidance of predator\cite{Stocker09} but it may also be due to the limitation in producing steering motion with only two flagella.
\noindent \textit{Volvox} on the other hand is a large multicellular organism which carries photoreceptors and thousands of flagella ($r = 500 \, \mu \mbox{m}$, $D_r = 2 \times 10^{-8} \mbox{ s}^{-1}$, $\Omega=1 \mbox{ s}^{-1}$, $D_r / \Omega = 2 \times 10^{-8}$). Our theory gives $\alpha^* \sim 0.0$ implying that continuous steering is the optimal for phototaxis of \textit{Volvox}. In fact, \textit{Volvox} coordinates thousands of flagella to make steering motion, and even has an adaptation mechanism.\\
Finally, we consider two limiting cases. First, in the limit of $\Omega \to \infty$, i.e. if steering accompanies no time cost, we have checked that $\mbox{Pe}$ monotonously decreases in $[0,\pi/2]$ so that the optimal angle is always $\alpha^*=0$. This corresponds to a ``perfect'' steering case, where the particle swims straight to the target to minimize the travel distance. The second case we want to consider is a ``passive'' reorientation: when $\phi \ge \alpha$, we now wait for the rotational diffusion to reorient it. In that case, the average reorientation duration $\langle \Delta t_{R} \rangle$ is given by
\begin{equation}
\langle \Delta t_{R} \rangle = \frac{2 \pi^2 + \alpha^2 + \alpha_R^2 -2 \pi (\alpha + \alpha_R)}{2 D_r}.
\end{equation}
Using this new equation for $\langle \Delta t_{R} \rangle$, $\mbox{Pe}$ becomes monotonously increasing in $[0,\pi/2]$ so that $\alpha^* = \pi /2 = 90^{\circ}$. 

\section{Discussion}

The feedback-control method presented in this paper mimics the ``run-and-tumbling'' of \textit{Escherichia coli} but combined with active steering, thus it is a simple but more efficient method to transport microscopic swimmers under thermal fluctuations. The added ability of deterministic ``active'' reorientation achieves more efficient transportation of particle than the natural ``run-and-tumbling''. It also has several advantages over other conventionally used micro-manipulation techniques: laser tweezers for example uses high power laser that could damage fragile samples and can be tricky to use as particles often jump out of the confining potential\cite{Merkt06}.\\
We have also addressed the problem of optimizing the feedback and observed some interesting insights. The active reorientation decouples the ABM and the reorientation process, in contrast to passive reorientation. Due to this decoupling, we showed that the optimal acceptance angle is a function of $D_r/\Omega$. Remarkably, since the timescale of passive reorientation is determined by $D_r^{-1}$, which scales in the cubic order of the radius, our method becomes particularly effective when the particles are relatively large. For instance in the case of a particle of diameter $3 \mbox{ } \mu \mbox{m}$, $D_r^{-1} \approx 18 \mbox{ s}$ in water whereas it can be as small as $\Omega^{-1} \approx 0.1$ s in our experiments, as shown in Fig.~\ref{U0_Omega}, leading to more than 10 times enhancement of Peclet number. For even smaller particles, the gain of active reorientation becomes less significant as $D_r$ approaches $\Omega$, though the magnitude of $\Omega$ is tunable by the applied electric field.\\
Although the optimal tolerance angle is determined by $D_r/\Omega$, our theory also predicts the robustness of the proposed algorithm. This is guaranteed by the fact that $\mbox{Pe}$ has a quite shallow shoulder towards large $\alpha$, at least for the range of parameters relevant to this experiment. In real world, individual particle may possess variable $\Omega$ or experience an inhomogeneous $D_r$ from the environment. It may also be possible that the exact position of the target might not be known, or estimation of the orientation might not be precise, contributing to a poor resolution of $\theta$. The final P\'eclet number however, is weakly affected by these noises thanks to the broad tolerance of optimal $\alpha$. That allowed us to control the motion of particles in spite of the strong variability in their behavior.\\
The key of our method lies in our finding of the peculiar rotational motion of Janus particle that can be switched on and off by changing the parameters of the electric field. Although the mechanism of this rotation is not yet fully understood, experimental measurements of $\Omega$ showed that it was proportional to $E_0^2$, implying that the torque $M$ as well as the force $F$ may originate from an asymmetric flow field around the particle generated by ICEO. The other parameter, $\omega$, has been poorly explored in the framework of ICEP. Further experimental studies, as well as theoretical works, should be addressed. Our results also suggested that geometrical factors such as chirality of the particle can play an critical role in determining the swimming behavior of the particle. An interesting challenge would be to find a way to artificially fabricate ``Brahma particles'', which as the Hindu god would have four ``heads''\cite{Maeda12}. These swimmers would have two well-designed axis of asymmetry, one used to propel the particle and the other to induce ``switchable'' rotations. Our experiment provides the first proof-of-principle demonstration of such an idea. Our results encourage further quest towards engineering functional artificial swimmers. \\ \hspace{1cm}

 \acknowledgments
The authors wish to thank Kyogo Kawaguchi and Daiki Nishiguchi for helpful discussions. This work was supported by JSPS KAKENHI Grant Number 12F02327.

\clearpage
\newpage

\part*{Materials and methods}
\subsection{Making of Janus particles} We used polystyrene spheres of diameter \scalebox{0.75}{$d = 3 \, \mu$}m. A droplet of a solution of theses polystyrene spheres is then dragged on a glass slide by a linear motor at the appropriate speed to obtain a monolayer of particles\cite{Prevo04}. We do not need a perfect crystal in our case but it is important that there is no particle on top of each other. Using thermal evaporation, one of their hemispheres is then coated by thin layers of chromium and gold with \scalebox{0.75}{$h_{Cr} \approx 10$} nm and \scalebox{0.75}{$h_{Au} \approx 20$} nm. The other hemisphere facing the glass slide remains bare polystyrene so that the particles have two hemispheres with different polarizabilities. After the coating process, the particles are detached from their substrate using mild-sonification and suspended in ion-exchange water. The observation of the particles at high magnification shows that they are almost always chiral (as shown in the inset of Fig.~\ref{scheme_Janus}), which is due to a fast slightly slanted evaporation process. Note that the amount of Janus particles exhibiting rotations at low field amplitude \scalebox{0.75}{$E_0$} increases when the metal layers is quite thick. For example, particles with \scalebox{0.75}{$h_{Cr} = h_{Au} \approx 10$} nm rarely exhibit rotations. Making chiral Janus particles thus requires to deposit large enough quantities of metal.

\subsection{ITO Electrodes} A droplet of a suspension of Janus particles is then put in between two ITO electrodes sandwiching a spacer made of stretched Parafilm of about \scalebox{0.75}{$40 \, \, \mu$}m. Using a function generator connected to the ITO slides, we apply a vertical AC electric field \scalebox{0.75}{$\boldsymbol{E}$} to the solution such that \scalebox{0.75}{$\boldsymbol{E} = E_0 \sin(\omega t) \, \boldsymbol{\hat{e}_z}$}. To prevent the particles from sticking to the bottom electrode, we apply a surface treatment to the ITO glass slides: the slides are exposed to a strong plasma for several minutes and are then immersed in a 5 \% solution of Pluronic F-127 (a non-ionic copolymer surfactant) for more than one hour. They are then washed with water to remove the excess of Pluronic. By coating the surface of the electrodes with this surfactant, we significantly reduce the risks of adhesion. However, this problem still remains one of the biggest experimental difficulties.

\subsection{Tracking and feedback} Throughout the experiment, particles were imaged using 10X, NA=0.3 objective lens mounted on a standard inverted microscope. Particles were illuminated with an incoherent light source, and the transmitted light was captured using CCD camera with 512x512 pixels at the frame rate of 100 fps. Tracking and feedback was done by a home-built LabVIEW program. Real-time tracking was initiated manually by feeding the program with the position of the particle of interest. Then for subsequent frames captured by camera, the small ROI around the target particle was extracted, then thresholded to obtain a binary image. The center of the mass was calculated from this binary image, and the updated particle coordinate was passed down to the next acquisition loop. At the same time, coordinate information was sent to the feedback loop, where it calculated the angle \scalebox{0.75}{$\theta$}. Feedback loop directly communicates with the function generator via USB connection, and updated the appropriate control parameters \scalebox{0.75}{$(\omega, E_0)$}.

\clearpage
\newpage

\part*{Langevin description and autocorrelation function}

The particles move at a constant speed $U_0$, rotate at the frequency $\Omega$ and are subjected to translational and rotational noises $\boldsymbol{\xi}_t$and $\xi_r$ respectively. If we assume that their motion is overdamped, we can thus write the following system of Langevin equations:
\begin{equation}
\left\{
\begin{array}{ll}
\dot{\boldsymbol{r}} = \boldsymbol{U_0} + D_t \boldsymbol{\xi}_t \\
\dot{\phi} = \Omega + D_r \xi_r.
\end{array}
\right.
\label{Lang_sys}
\end{equation}
\begin{figure}
\begin{tabular}{cc}
\centerline{\includegraphics[width=4 cm]{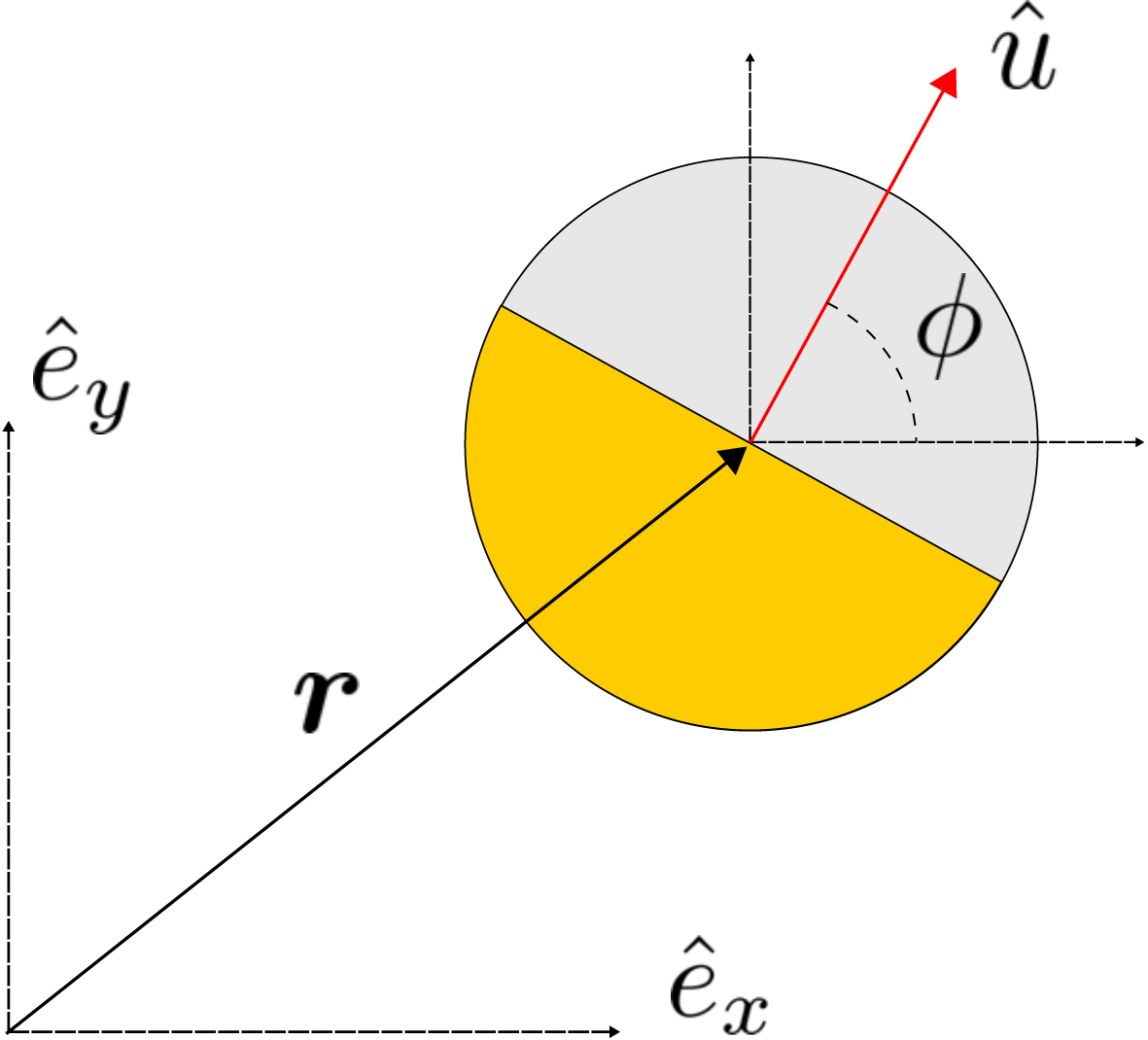}}
\end{tabular}
\caption{\label{scheme}Two-dimensional scheme of a Janus particle}
\end{figure}
This model had already been studied to get an analytical expression for the mean-square displacement of L-shaped artificial swimmers\cite{Kummel13} but here, we will focus on the autocorrelation function instead. The second equation can easily be integrated to get
\begin{equation}
\phi(t) = \Omega t + D_r  \int^{t}_{0} \xi_r(t') dt' + \phi_0.
\label{gaussian}
\end{equation}
$\xi_r$ being a white noise, we have
\begin{equation}
\begin{array}{cl}
\langle \phi(t) \rangle & = \phi_0 + \Omega t \\
\langle \left(\phi(t) - \langle \phi(t) \rangle \right)^2 \rangle & = 2 D_r t.
\end{array}
\label{phi}
\end{equation}
According to Eq.~[\ref{gaussian}], $\phi(t)$ is a sum of gaussian variables and is therefore a gaussian itself. As we just calculated its first and second moments, we can deduce the expression of the probability density
\begin{equation}
P(\phi,t)  = \frac{1}{\sqrt{4 \pi D_r t}} \exp \left( - \frac{(\phi - \phi_0 - \Omega t)^2}{4 D_r t}\right),
\label{dens_prob}
\end{equation} 
and the Green function
\begin{equation}
G(\phi_1,\phi_2,t_1 - t_2)  = \frac{1}{\sqrt{4 \pi D_r (t_1 - t_2)}} \exp \left( - \frac{(\phi_1 - \phi_2)^2}{4 D_r (t_1 - t_2)}\right).
\label{Green}
\end{equation} 
\begin{figure}
\centerline{\includegraphics[width=7 cm]{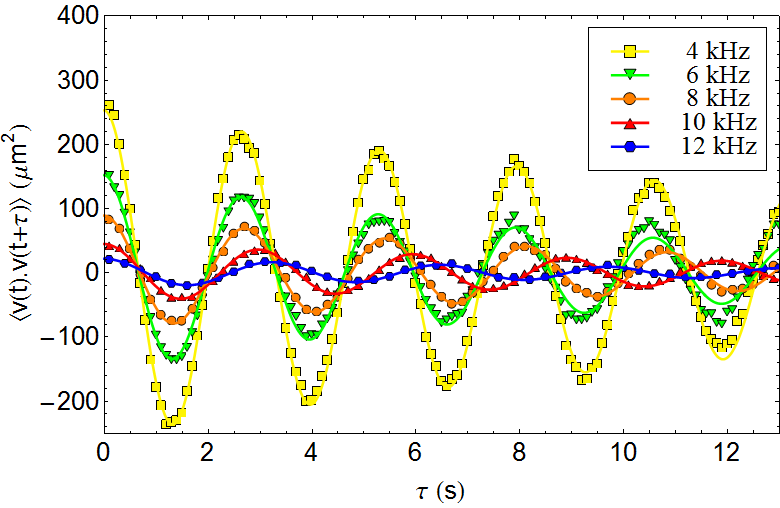}}
\caption{\label{correlations}Time evolution of the auto-correlation function for different frequencies $\omega$. The symbols are experimental measurements and the lines of the same colors correspond to a fit using expression~\ref{vcorr}. We have three fitting parameters: $\Omega$, $U_0$ and $D_r$.}
\end{figure}
The velocity autocorrelation function of a particle at time $t$ is given by
\begin{equation}
\begin{array}{l}
\langle \boldsymbol{v}(t) \cdot \boldsymbol{v}(t+\tau) \rangle \\ \\= \left\langle \left[ U_0 \boldsymbol{\hat{u}}(t) + 
D_t \boldsymbol{\xi}_t(t) \right] \cdot \left[ U_0 \boldsymbol{\hat{u}}(t+\tau) + D_t \boldsymbol{\xi}_t(t+\tau) \right] \right\rangle\\
\\
 = 4 D_t \delta(\tau) + U_0^2 \langle \boldsymbol{\hat{u}}(t) \cdot \boldsymbol{\hat{u}}(t+\tau) \rangle \\ \\
 = 4 D_t \delta(\tau) + U_0^2 \left\langle \left( 
\begin{array}{l}
\cos\phi(t)\\
\sin\phi(t)
\end{array}
\right)
\cdot \left( 
\begin{array}{l}
\cos\phi(t+\tau)\\
\sin\phi(t+\tau)
\end{array}
\right)
\right\rangle \\ \\
 = 4 D_t \delta(\tau) + U_0^2 \left\langle \cos\phi(t)\cos\phi(t+\tau) + \right. \\ \\
 \hspace{0.5 cm}\left. \sin\phi(t)\sin\phi(t+\tau) \right\rangle.
\end{array}
\label{averages}
\end{equation}
Here, $\langle ... \rangle$ represents an ensemble average given, for an arbitrary function $f \left( \phi_1, \phi_2,t_1 - t_2 \right)$ by
\begin{equation}
\begin{split}
\langle f \left( \phi_1, \phi_2,t_1 - t_2 \right) \rangle = \int^{\infty}_{-\infty} \int^{\infty}_{-\infty} f \left( \phi_1, \phi_2,t_1 - t_2 \right) P(\phi_2,t) \\
G(\phi_1,\phi_2,t_1 - t_2) \, d\phi_1 \, d\phi_2.
\end{split}
\label{ensemble}
\end{equation}
Injecting Eqs.~[\ref{dens_prob}] and~[\ref{Green}] into~[\ref{ensemble}], we can calculate the two averages of Eq.~[\ref{averages}] and find as a final expression for the velocity auto-correlation function
\begin{equation}
\langle \boldsymbol{v}(t) \cdot \boldsymbol{v}(t+\tau)\rangle = 4 D_t \delta(\tau) + U_0^2 \,\exp(-D_r \tau) \, \cos(\Omega \, \tau).
\end{equation}
Strictly speaking, this expression diverges at $\tau=0$ but the system of Langevin Eqs.~[\ref{Lang_sys}] is actually only valid at times greater than the typical collision time of the heat bath. We can thus neglect the first term. If we use the expressions of the average velocity of the particles $U_0$ and of the angular velocity $\Omega$, we recover Eq.~[\ref{vcorr}]. The agreement with the experimental results is excellent as can be seen on Fig.~\ref{correlations}.

%\bibliographystyle{siam}
%\bibliography{bugsrefs}
%%\bibliographystyle{elsart-num} revtex4 will call the correct apsrev4-1.bst file.

\end{document}